# Localization for Anchoritic Sensor Networks

Yuliy Baryshnikov ◇      Jian Tan †

◇Bell Laboratories, Murray Hill, 07974 NJ
†EE Department, Columbia University, New York, 10025 NY

*Abstract*— We introduce a class of *anchoritic* sensor networks, where communications between sensor nodes is undesirable or infeasible, due to, e.g., harsh environment, energy constraints, or security considerations.

Instead, we assume that the sensors buffer the measurements of the physical phenomena over the lifetime, and report measurements directly to a sink (note that "report" does not necessarily require direct communication). Upon retrieval of the reports, all sensor data measurements will be available to a central entity for post processing.

Our algorithm is based on the further assumption that some of the data fields that are being observed by the sensors can be modeled as a local (i.e. having decaying spatial correlations) stochastic process; if not, then choose another auxiliary field, such as hydroacoustic noise, cloud shadows cast on the ground, or animal heat. The sensor nodes record the measurements, or a function of the measurements, e.g., "1" when the measured signal is above a threshold, or "0" otherwise. These sensors generated time-stamped sequences are ultimately transferred to the sink. The localization problem is then approached by analyzing the correlations between these sequences at pairs of nodes.

As engineering ramifications to the idea of anchoritic sensors, we discuss the localization scheme for sensors deployed on the seabed, where radio signals are strongly attenuated in sea water within feet of their transmission; also, we discuss a two-tiered architecture tasked with combining heterogenous nodes: deaf sensors and local masters.

## I. Introduction

The "coordinate-free" localization problem in sensor network has attracted significant attention in the literature, see, e.g. the survey [1]. The problem is to determine positions of the nodes in the network without using absolute reference information, like GPS or direction/distance information relative to some known beacons. Coordinate-free localization problem is therefore to determine the *absolute* positions of the nodes using only the *local information*, e.g., the internode distances or relative directions. This local-to-global localization problem presents a serious research challenge and the amount of work on it is rapidly growing, see, e.g., [2]–[15]. However, in majority of publications, addressing the localization problem either assumes extensive *internode communications*, i.e., a bidirectional exchange of signals which is used to infer the pairwise distances; or assumes some variant of a *system of beacons* that can send globally structured calibrating signals (*"poor man's GPS"*).

Within the framework of internode communications, one uses one or a combination of RSS (received signal strength), ToA (time of arrival) and AoA (angle of arrival) data to reconstruct the mutual positions of the nodes, and consequently to determine their absolute positions. The beacons with known positions provide absolute reference points for the remaining sensor nodes.

Approaches with globally broadcasted and centrally steered signals that reach a significant fraction of all the sensors can also be efficiently used to calibrate locations. For example, the Spotlight system [16], with the aid of steerable laser rays that sweep over the monitored terrain obtains the locations of sensor nodes without equipping them with specialized ranging hardware. We note however that this requires precise knowledge on the trajectory of the ray.

In this paper, we look into the networks subject to severe communication constraints. In particular, we do not allow either *internode communications* nor any *centrally structured signals* to be used for the localization.

We decouple the localization problem into two steps: the first stage is to recover the internode mutual positions, and the second stage is to use the information obtained from the first stage to reconstruct the global positions of the nodes.

Traditionally, the localization problem (in the absence of centrally steered signals) implicitly assumes that the relative position information should be gained from the signal exchange (*chatter*) between the nodes, and that the subsequent processing of the exchanged signals is necessary. However, as we argue below, the requirement that the sensors are able to regularly emit signals and process information is undesirable in many applied situations, due to, e.g., harsh environment, energy constraints, or security reasons. Assuming this for an instant, we ask:

> *can the localization problem, or rather the first stage of the localization problem — recovery of the internode distances, be solved under the conditions that the sensor nodes do* not *have the ability to exchange signals, and that no global signals are applicable for localization?*

Clearly, the sensors have to be able to gather some measurements and eventually report them to a sink/processing entity; a sensor which is unable to do even that much can be removed without any detrimental effect for the network operation.

We will refer to sensor networks subject to constraints of these types (deprived of the ability to chat[1] and lacking centrally controlled broadcasted signals) as *anchoritic* sensors.

---
[1]To be more precise, chatting is a two-way process involving listen and talk, while for anchoritic sensors, either the sensors can not listen or they can not talk, or even neither.

It is perhaps counterintuitive that the localization problem for anchoritic sensor networks can be solved. Before presenting our approach, we need, however, answer the following natural questions:
1) When and where anchoritic sensor networks are necessary?
2) If no communication allowed, how can the measurements collected by each individual sensors be transferred to the sink/center?

These questions are addressed in the next subsection. In the last subsection of the introduction we describe related work. The rest of this paper is organized as follows: after describing the essentials of our approach to the localization problem for anchoritic sensor networks and some engineering ramifications, we give several models of the environmental random fields in section III. These models are candidates for recovering the mutual distance data in anchoritic sensor networks. Section V presents the results of our numeric simulations, which are followed by the conclusion.

*A. Motivation*

*1) Anchoritic sensors:* Situations where the sensors collecting data are anchoritic are much more widespread than one might suppose.

First of all, when the sensor networks are immersed into a harsh environment where the communications between the sensor nodes is infeasible, standard approach that uses internode communication signals to infer the pairwise distances no longer applies. As an important application example, we consider the sensors deployed on the seabed. These sensors would have difficulties using the radio signals to communicate, because electromagnetic waves are strongly attenuated in water. Even using alternative underwater acoustic signals, for a large scaled sensor network, is not a good solution, since, compared with electromagnetic waves, the latency using acoustic signals is typically much higher as sound waves travel much slower , and, due to the multi-path propagation and noise characteristics, the effective data rates and packet loss is much greater as well [17]. Hence one might resort to anchoritic assumptions for large scaled underwater sensor networks.

Also, if the number of sensors grows to $10^5$-$10^6$, the cost of any extra feature with which the sensors are provided scales correspondingly. Though requiring the ability to listen and to analyze the received signals increases the cost of a single device by only an innocuous amount of expense, it results in a prohibitively expensive cost for a large sensor network. Hence, the option of only choosing *mute or deaf* sensors that are cheap, or perhaps to combine emaciated sensors with expensive and powerful ones, might force the developers to adopt the *anchoritic* requirements.

Another obvious situation where the sensor chatter is undesirable is the battlefield or other applications involving an adversary. Transmitting signals during the mission reveals the presence of the sensors, and therefore makes them vulnerable to suppression and manipulation. Similarly, one would not deploy any globally structured signals for localization purposes as an adversary could generate noise or worse, emulate the system signals to compromise the localization completely.

While there are further scenarios where anchortic networks could be necessary, these situations — physical obstacles, cost considerations, security requirements — seem to cover most of them.

*2) Information delivery:* Now, if the sensors in the network are silent, how can they report the data to the sink? There are again several scenarios.

First, the sensors (or their data storage units) can be indirectly collected after their mission is completed. As for the sensor networks deployed on the seabed. Buoyant sensors can be attached to heavy ballasts and sink themselves. After a period of time during which the sensors perform their measurements, the ballasts are released and the sensors emerge to the surface where their measurements can be collected either manually or by radio signals. Clearly, the original positions of the sensors cannot be reliably estimated just by their locations on the sea surface.

Another case is relevant to those sensors which can only report their collected measurements within a short period of time after being activated by extraneous command signals and have to keep silent otherwise. For example, sensors serving for military or security purposes in some special applications should not reveal themselves until the surroundings are safe. Here, the sensors are equipped with radios capable of talking. However, for the reasons stated above, the communication should be on demand and short, e.g., sensors deployed in the enemy's rear area. Hence, the sensors transmit just once during their operation cycle, transferring to the center (which can be, e.g., a mobile agent that passes by) all the information they gathered. After that, the sensors either wait for another transmission cycle, or even are compromised.

In either case, the central entity should be able to recover the original positions of the sensors depending only on the individual measurements by each sensor, and this information is oblivious of the positions and the existence of other sensors in the network.

*B. Related Work*

As we mentioned above, most of the publications on the "coordinate-free" localization problem follow the path of reconstructing locations from the proximity data. These approaches typically assume the distances being given by RSS data, or just by the connectivity patters of the network formed by the sensors. Then some geometric considerations are used, followed possibly by iterative adjustment and fine tuning. The works following to some degree this pattern are, e.g., [2]–[15].

Some deviations from this scheme are also considered in literature. For example, the work [18] assumes a lack of communications between the nodes, yet relies on several anchors which can communicate with significant parts of the network. The distances to these anchors are then used for the localization.

Similarly, [19] assumes a system of beacons having known positions and sending acoustic signals used for the localization.

A somewhat more complicated approach mixing internode chatter and beacons is used in [20].

The localization system [16] achieves high accuracy in recovering coordinates of the nodes without requiring internode chatter. However, some ranging signals (steerable laser rays sweeping over the terrain populated with sensors) are necessary.

In the existing literature, the one most close to our techniques appears to be the ingenious SLAT (simultaneous localization and tracking) proposal [21]. There, the authors consider a network of cameras tracking a moving object by recovering their own positions and then the trajectory of the tracked object. While conceptually not completely disjoint from our model of random walkers (see Section III-C), the approach of [21] relies heavily on the uniqueness of the moving object and on the far range of their sensing devices. Introducing many targets seems to require a major overhaul of the approach used there, which might lead to a statistical procedure of distinguishing between multiple targets, and thus to techniques close to ours.

On a more conceptual level, the correlations between the measurements have been used in the sensor/ad-hoc networks, most notably to develop coding schemes. It has been proposed to use correlations in the measurements to improve the network throughput. Similarly, using correlated signals in ad-hoc wireless MIMO networks can improve the transmission rates. Here, however, we do not try to filter the noise out of the signals, but rather to use the noise (insofar it admits some decaying correlation functions) for the localization.

## II. Approach

We approach the problem of recovering the mutual distances between the nodes in an anchoritic sensor network by exploiting the time-space correlation structure of some random field observed by the sensors. This random field can be what the sensors are tasked with measuring, or some auxiliary measurements performed with the sole purpose to solve the localization problem.

### A. Description and Definitions

Our algorithm is based on the assumption that the primary data field observed by the sensors can be modeled as a locally isotropic random field (for a precise definition, please refer to the discussion preceding Theorem 1); if not, then another auxiliary physical phenomenon, which is locally stationary, can be alternatively measured by the sensor nodes, e.g., hydroacoustic noise near the seabed, cloud shadows cast on the ground, or even artificially introducing auxiliary signals like broadcasting random radio noise mandatorily. The sensor nodes record the measurements, or a function of the measurements, e.g., simply recording "1" when the measured signal is above a threshold, otherwise record "0". Therefore, only sequences of $0, 1$ bits as well as data measurements will be transferred to the sink. The crucial intuitive idea behind this approach is the correlations between these sequences at different sensors decrease with the distance between the sensors. Then, by analyzing the correlation between these boolean sequences at pairs of nodes, we can approach the localization problem indirectly. Notice that, though each sensor has to reserve a storage space for recording the sequence of 0's and 1's, this requirement only increases a negligible overhead, since even only 250 bytes contains 2000 bits.

More precisely, we assume that a locally stationary random field (defined by conditions in Theorem 1) $\xi$ is measured by the sensors $\mathbf{N} = \{1, 2, \ldots, N\}$ in the network at synchronized instants $t_o, t_1, \ldots, t_T$. The position of sensor $i$ is denoted as $z_i = (x_i, y_i)^2$. The sensor nodes record "1" when the measured signal is above a threshold, otherwise record "0" (this can be generalized to a multivalued record).

*Definition 1:* For a subset $I = \{i_1, i_2, \ldots, i_S\} \subset \mathbf{N}$, define empirical instantaneous correlation functions to be

$$\kappa^T(I) = \frac{1}{T} \sum_{j=1}^{T} \xi_{i_1}(t_j) \xi_{i_2}(t_j) \cdots \xi_{i_S}(t_j), \qquad (1)$$

where $\xi_i(t_j)$ is the record of the field $\xi$ by the sensor $i$ at instant $t_j$.

*Remark 1:* Here the requirement for synchronization is not tight, since even when clocks drift over a long period of time, a small time lag between the measurements will not affect the space-time correlation too much.

*Remark 2:* In this paper, we are only interested in analyzing the correlation at pairs of nodes, and therefore $S = 2$. For $S \geq 3$, it contains more information on the relative positions of the sensors (e.g., $S = 3$ forms a triangle). We defer this discussion to later work.

### B. Theoretic Framework

When the field $\xi$ is stationary and ergodic with respect to time $t$, the empirical correlation function, in the limit of $T \to \infty$, converges to its expected value (e.g., see Theorem 9.6 of [22]),

$$\kappa(I) = \mathbb{E}\xi_{i_1}\xi_{i_1}\cdots\xi_{i_S}. \qquad (2)$$

In many cases, when the sensors form spacially separated clusters, $I_\alpha, \cup I_\alpha = I$, the correlation function also clusters correspondingly to be

$$\kappa(I) \approx \prod_{I_i \subset I} \kappa(I_i), \qquad (3)$$

where the approximation is good when the distances between the points of different $I_i$'s are large.

Correlation functions give rise to cumulants. Cumulants $\{c_l\}_{l \geq 1}$ are defined as satisfying

$$\sum_l \frac{c_l s^l}{l!} = \log(\sum_l \frac{\kappa_l s^l}{l!}), \qquad (4)$$

where $\kappa_l = \kappa(\{1, \ldots, l\}))$ are the correlations. The cumulant functions are nonvanishing at the diagonal (that is for the sets of points spatially close) and vanishing far from the diagonal (that is where the pairwise distances between the points grow), e.g., see [23].

---

[2]We concentrate on the plane localization problem in this note; our approach can be used in a high dimensional space or generally, an abstract metric space. For example, the sensors can be viruses planted at unknown nodes of a virtual networks unseen from behind a firewall.

We will concentrate on the pairwise cumulants, which reduce to the standard statistical correlation

$$c_2(i,j) = \mathbb{E}\xi_i\xi_j - \mathbb{E}\xi_i\mathbb{E}\xi_j, \quad (5)$$

for sensor $i$ and $j$. The notation $c(z_i, z_j) \equiv c_2(i,j)$ may also be used when the exact locations of sensor $i$ and $j$ need to be stated.

*Remark 3:* Notice that $c(i,i) = \mathbb{V}ar(\xi_i) > 0$ and also $c(i,i) \geq |c(i,j)|, z_j \neq z_i$. Further, if the correlation is continuous function of $z_i, z_j$, $c(i,j) > 0$ for sensor $i$ and sensor $j$ close enough.

The cumulants described above capture the dependence between the *instantaneous* values of random fields. One can also exploit also the dependence between different time values at different points. A general way to do so is to extend the dimension of the random field $\xi$. Consider the new field $\tilde{\xi}_\Delta$ whose value at point $z$ and time $t$ is the trajectory of $\xi$ between $t - \Delta$ and $t + \Delta$:

$$\tilde{\xi}(z,t) = \{\xi(z,s)\}_{s\in[t-\Delta,t+\Delta]}. \quad (6)$$

The 2-point correlation functions involve in this case some general kernel functions $K(s,s')$ and are given by

$$\langle \tilde{\xi}, \tilde{\xi}' \rangle = \int_{[-\Delta,\Delta]^2} \xi(s)K(s,s')\xi(s')dsds'. \quad (7)$$

In Section III we will present several examples where the values of empirical cumulants can serve as an efficient proxy for the internode distances, thus effectively resolving the first stage of the localization problem for the anchorite sensor networks.

*C. Proximity graph*

To solve the localization problem, we need construct the proximity graph. We assume a standard setting: in a plane region $A \subset \mathbb{R}^2$, $N$ points are selected independently and uniformly with respect to Lebesgue measure. We will assume throughout the note that the area of $A$ is normalized to the unity.

We will reconstruct the positions of the nodes with respect to the first $B < N$ points (called local *beacons*), whose coordinates are assumed to be known[3]. The positions of the rest sensors are unknown, and are to be determined.

If the mutual distances $d_{ij} = |z_i - z_j|$ are known for all pairs of sensors, it is possible to reconstruct the whole configuration $\mathcal{Z} = \{z_1, \ldots, z_N\}$ up to an isometry of the plane preserving the positions of beacons (that is up to a rotation if $B = 1$ or up to an axis symmetry if $B = 2$). Obviously, the positions and the number of beacons have impact on the reconstruction.

In an anchoritic network, the sensors do not know their mutual displacements, and we resort to the measurements of (empirical) correlations/cumulants as a proxy for the internodes distances.

Denote the empirical correlation function of sensor $i$ and sensor $j$ by $c(z_i, z_j)$, and empirical correlation by $c^{(T)}(z_i, z_j)$. *We will make the assumption that, with the sample size large enough the cumulants $c(z_i, z_j)$ for all pairs $(i,j)$ can be approximated with desired precision by the empirical cumulants.*

To construct the proximity graph $\Gamma_N$ we connect each node $i$ to $k_N$ nodes having the *largest* values of the empirical cumulants. For an ergodic field such that

$$c(z,z) > c(z,z'), z \neq z',$$

and $k_N = o(N)$, the resulting graph $\Gamma_N$ will approximate the corresponding $k_N$-nearest neighbor graph based on the Euclidean distances between nodes [4].

We need the cumulants to satisfy the following two assumptions.

A: Convergence of empirical cumulants: $c^{(T)}(z_i, z_j) \to c(z_i, z_j)$ for all $i, j \in \mathbf{N}$ as $T \to \infty$, and,

B: Asymptotic isotropy: for all $x, y \in A$, the set $S_x(\delta) \triangleq \{y : c(x,y) \geq c(x) - \delta\}$ satisfies

$$\frac{S_x(\delta)\sqrt{\pi}}{\sqrt{\text{Area}(S_x(\delta))}} \to \text{a unit circle}, \quad (8)$$

as $\delta \to 0$.

*Remark 4:* Intuitively speaking, Assumption B means that the left hand side of (8) can be sandwiched between two circles with radii arbitrarily close to 1, as $\delta \to 0$.

Then , the following result is valid.

*Theorem 1:* Under Assumptions A and B, let $k_N = \log N^c, c > 1$, then, for any $i$,

$$h_{ib}^{(N)}\sqrt{\frac{k_N}{\pi N}} \to d_{ib}, \quad (9)$$

for any $b = 1 \ldots, B$, with high probability. Here $h_{ij}^{(N)}$ is the distance between nodes $i$ and $j$ in $\Gamma_N$ (hop distance) and $d_{ij}$ is the Euclidean distance between $z_i$ and $z_j$.

*Proof:* First, we need prove the following result. Connecting each pair of sensors with a distance less than

$$r(N) = \sqrt{\frac{(\log N)^c}{\pi N}}, \ c > 1, \quad (10)$$

one obtains a graph $G(N)$. Let $h_{ij}^{(G)}$ be the hop distance between nodes $i$ and $j$ in $G(N)$. Then, for any $\epsilon > 0$, we have

$$\lim_{N\to\infty} \mathbb{P}\left[\left|h_{ij}^{(G)}r(N) - d_{ij}\right| < \epsilon, \forall Z_i, Z_j \in \mathcal{Z}\right] = 1. \quad (11)$$

The argument goes as follows. Choose $\rho(N) = \sqrt{\frac{c_1 \log N}{\pi N}}$ where $c_1 > 2.5$. For sensor $i$ and $j$ with positions $Z_i$ and $Z_j$, connect them by a sequence of circles of radius $\rho(N)$ such that the centers of the adjacent circles are $r(N) - 2\rho(N)$ away (when $N$ is large, $r(N) - 2\rho(N) > 0$). We call sensor $i$ and

---
[3]This is not a departure from the assumptions of anchoritic nodes: these special sensors do not broadcast any specially structured signals.

[4]A $k_N$-nearest neighbor graph is build on a metric space in such a way that each vertex is connected to $k$ nearest vertices.

$j$ to be $\rho$-vicinity connected if there exists at least one sensor lying in each of the circles along the line $(Z_i, Z_j)$. This is shown in Figure 1.

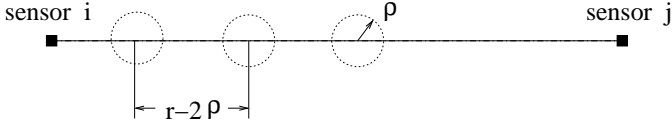

Fig. 1. $\rho$-vicinity connection

Therefore, for any $i, j \in \mathbf{N}$ and large $N$, we have

$$\mathbb{P}[\, i,j \text{ are not } \rho\text{-vicinity connected}]$$
$$= \mathbb{P}[\text{at least one circle along } (Z_i, Z_j) \text{ is empty}]$$
$$\leq \mathbb{E}[\text{number of empty circles along } (Z_i, Z_j)]$$
$$\leq \frac{\sqrt{2}}{r - 2\rho} \mathbb{P}[\text{one given circle is empty}]$$
$$= \frac{\sqrt{2}}{r - 2\rho}(1 - \pi\rho^2)^N$$
$$\sim \sqrt{\frac{2\pi N}{(\log N)^c}} N^{-c_1}$$
$$\leq N^{-(c_1 - \frac{1}{2})}, \quad (12)$$

implying,

$$\mathbb{P}[\, i,j \text{ are } \rho\text{-vicinity connected}, \forall i,j \in \mathbf{N}]$$
$$\geq 1 - \sum_{i \neq j \in \mathbf{N}} \mathbb{P}[\, i,j \text{ are not } \rho\text{-vicinity connected}\,]$$
$$\geq 1 - \binom{n}{2} n^{-c_1 + \frac{1}{2}}. \quad (13)$$

Passing $N$ to infinity, we obtain

$$\lim_{N \to \infty} \mathbb{P}[\, i,j \text{ are } \rho\text{-vicinity connected}, \forall i,j \in \mathbf{N}] = 1.$$

Hence

$$\lim_{N \to \infty} \mathbb{P}\left[ \left| h_{ij}^{(N)} r(N) - d_{ij} \right| < \epsilon, \forall Z_i, Z_j \in \mathcal{Z} \right]$$
$$\geq \lim_{N \to \infty} \mathbb{P}[\, i,j \text{ is } \rho\text{-vicinity connected}, \forall i,j \in \mathbf{N}]$$
$$= 1. \quad (14)$$

Next, choosing $r_1(N) = \sqrt{1-\epsilon} \cdot r(N)$ and $r_2(N) = \sqrt{1+\epsilon} \cdot r(N)$, we obtain graph $G^{-\epsilon}(N)$ and $G^{+\epsilon}(N)$, respectively. Actually, we can prove that $\mathbb{P}[G^{-\epsilon}(N) \subset \Gamma(N) \subset G^{\epsilon}(N)] \to 1$, when $N \to \infty$, and the argument goes as follows.

In graph $G^{-\epsilon}(N)$, we define for sensor $i$ the indicator function

$$X_{ij} = \mathbf{1}(\text{sensor } i \text{ and sensor } j \text{ are connected}). \quad (15)$$

For all $j \neq i$, $\{X_{ij}\}$ are i.i.d random variables. Let $Y_i$ be the number of neighbors in $G^{-\epsilon}(N)$ for sensor $i$. Then,

$$\mathbb{P}[Y_i > k_N] = \mathbb{P}\left[ \sum_{j \neq i} X_{ij} > k_N \right]$$
$$= \mathbb{P}\left[ \sum_{j \neq i} X_{ij} > \frac{N}{\sqrt{1-\epsilon}} \mathbb{E}[X_{ij}] \right], \quad (16)$$

and, from well-known large deviation results, there exists $\theta > 0$ such that

$$\mathbb{P}[Y_i > k_N] \leq e^{-\theta N}. \quad (17)$$

Therefore,

$$\mathbb{P}[Y_i > k_N, \text{for some } i \in \mathbf{N}] \leq \sum_{i=1}^{N} \mathbb{P}[Y_i > k_N]$$
$$\leq N \cdot e^{-\theta N}$$
$$\to 0 \text{ as } N \to \infty. \quad (18)$$

From (18) and condition (8), we have as $N \to \infty$,

$$\mathbb{P}[G^{-\epsilon}(N) \subset \Gamma(N)] \to 1. \quad (19)$$

By the same argument, as $N \to \infty$, we can also prove

$$\mathbb{P}[\Gamma(N) \subset G^{\epsilon}(N)] \to 1. \quad (20)$$

Combining (11), (19) and (20), then passing $\delta \to 0$, we complete the proof. ∎

From this theorem, we see that knowing the cumulants is enough to reconstruct, with arbitrary precision, the positions of all nodes in the networks, assuming that the network is large enough.

In the next section we consider several models for various "real-life" inspired ergodic random fields which could be used for the distance estimation purposes.

### III. MODELS OF RANDOM BACKGROUND FIELDS

In this section, the sensors are assumed scattered uniformly in an open space. For the case when the sensors are not uniformly distributed, one might need introduce local masters/sinks. We will discuss the ramifications in Section IV.

#### A. Boolean Model

This model imitates an anchoritic sensor network with the shadow/light patterns used as the auxiliary random field. The shadow/light detection can be done in an inexpensive fashion in terms of both hardware costs and energy consumption.

To model the shadow patterns generated by "clouds" (the clouds here can be real clouds, or artificially introduced acoustic noise signals or electromagnetic disturb), we will apply the widely used *Boolean model* (see e.g. [24]). The model is specified by a point process $\mathcal{P}$ and a class of bounded random sets $\mathcal{B}$. To keep matters simple, we assume that $\mathcal{P}$ is a Poisson point process, and the $B \in \mathcal{B}$ is a circle with a random radius $R$. Given the pair $(\mathcal{P}, \mathcal{B})$, the random set $C$ is

$$C = \bigcup_{Z_\alpha \in \mathcal{P}} (Z_\alpha + B_\alpha),$$

where $B_\alpha$ are *iid* realizations of the sets from $\mathcal{B}$.

At each instant $t$, the sensors located in the field observe

$$\xi(z) = \begin{cases} 1 & \text{if } z \in C \text{ and} \\ 0 & \text{otherwise.} \end{cases}$$

For thus constructed random field $\xi$, the correlation function between any 2 points $z, z'$ can be explicitly computed. Let $\mathbb{P}(d\beta)$ be the probability measure on the shape space $\mathcal{B}$. For any $\beta \in \mathcal{B}$ and $z \in \mathbb{R}^2$ define

$$\psi(z) = |\beta \cap (\beta + z)|$$

to be the Lebesgue measure of the intersection of *shape* and its displacement by $z$. For example, if *shape* is a ball of radius $r$, then, for any $|z| \leq 2r$,

$$\psi(z) = 2\left(r \arccos \frac{|z|}{2r} - \frac{|z|\sqrt{4r^2 - |z|^2}}{4}\right), \quad (21)$$

and, therefore, one has

*Proposition 1:*

$$c(z, z') = \mathbb{E}\xi(z)\mathbb{E}\xi(z')\left(\int_\mathcal{B} e^{-\psi(z-z')}\mathbb{P}(d\beta) - 1\right)$$

and

$$\mathbb{E}\xi(z) = \int_\mathcal{B} e^{-|\beta|}\mathbb{P}(d\beta).$$

If the time intervals between each measurement are spaced far enough, one can model the realizations of the random sets $C_t, t = 1, \ldots, T$ as *iid*, and the empirical correlation function between any two points $z, z'$ converges to $c(z, z')$. Applying formulae of Proposition 1 one can recover the internode distance data.

### B. Large Clouds

A variant of the Boolean model deals with the unbounded shapes, the "large clouds". Here the clouds are represented as the parallel strips of random widths. More precisely, we consider the random set $C$ to be bounded by a family of parallel lines, which are orthogonal to a direction that is chosen uniformly from the unit circle, and whose orthogonal projections to this direction form a Poisson point process of some constant intensity (one can check that this definition is independent of the choice of the origin in the plane).

In other words, $z \in C$ if

$$x_{2i} \leq \langle z, e \rangle \leq x_{2i+1}, i = \ldots, -1, 0, 1, \ldots,$$

where $\langle, \rangle$ denote Euclidean scalar product, $e$ is a random vector chosen uniformly from $\{|e| = 1\}$ and $\{x_k\}_{-\infty}^{\infty}$ is a Poisson point process with constant intensity $\lambda$.

Again, in this situation it is easy to find the 2-point cumulant. In fact, if a random set is translation and rotation invariant, and each realization of the set is a union of infinite strips, then

*Proposition 2:*

$$c(z, z') = a - b|z - z'|$$

for some constants $a, b > 0$.

Once again, we see that the cumulant attains its maximum on the diagonal, and therefore can be used for recovering the distance data in anchoritic sensor networks.

### C. Random Walkers

This model describes the random field generated by some independently moving objects: as an example, one can imagine sensors registering heat/verberation of an animal moving nearby.

To model this, we consider random field represented by a family of random walkers in the area $A$. At time $t$ a sensor located at position $z$ registers $\xi = 1$ if there is a walker at a distance at most $r$ from $z$.

We will use the spatio-temporary version of the correlation function, that is we will consider also the lagged correlations

$$\sum_s \mathbb{E}\xi(z,t)\xi(z',t+s), |s| < \Delta. \quad (22)$$

This trick stretches the cumulants of close points over the interval $\Delta$. The precise expression for the cumulants in this model is a polynomial in Gaussian functions and is quite cumbersome, so we will not present it here. We just state

*Proposition 3:* The 2-point cumulant function for the random walkers model is positive on the diagonal and vainishing at the infinity.

## IV. ENGINEERING RAMIFICATIONS

The idea of anchoritic sensors can be extended to situations where heterogenous sensors are combined. Some sensors are powerful to chat and can form the backbone of local masters/sinks, while other sensors are so weak that they can only collect measurements and report them to a nearest sink. The network architecture is depicted as in Figure 2.

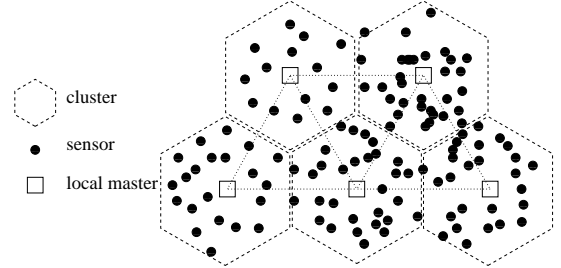

Fig. 2. Combine Deaf Sensors and Local Sinks

The sinks are powerful nodes that can exchange information and, if possible, can even engineered to be able to generate radio or acoustic signals at random time with random strength. These auxiliary signals will be measured by other emaciated sensors. This measuring procedure can be simply record 0 or 1 by judging the strengths of the signals.

Sensors in a cluster periodically report their data to the local master/sink. Using the correlation values, the local master can estimate the positions of the sensors in its cluster. For this structure, the sensors are only required to be uniformly scattered within a cluster, and the density can change for different clusters.

Once again, the distinguishing feature of this approach is that it does not require the broadcasting of globally structured signals from a central entity.

## V. EXPERIMENTAL RESULTS

In this section we present simulation experiments on proximity graph reconstruction in anchoritic networks, based on several random fields described above. In all these simulations, $N = 1000$ sensors were chosen independently at random from uniform density in the unit square $A$, and the simulation goes in a discrete fashion $t = 1, 2, \cdots, 2000$. These sensors are represented as little squares on the corresponding plots. The proximity graph $\Gamma$ is formed by connecting a given number $\left(k_N = \lfloor (\log N)^{1.2} \rfloor = 10\right)$ of pairs of nodes with the largest value of the empirical cumulants (the reason why we choose $k_N = (\log N)^{1.2}$ comes from the proof of Theorem 1).

### A. Round clouds — Boolean model

In this simulation, round clouds of random radii uniformly distributed on $[0, .2]$ are modeled as a Poisson random field with density 30. Blue circles on Figure 3 depict a realization of the Boolean model; the empirical cumulants were formed from $T = 2000$ independent samples.

The proximity graph $\Gamma$ shown on Figure 3 is formed by connecting a given number ($k_N = 10$) of pairs of nodes with the largest value of the empirical cumulants. one can see that this graph strongly resemble a nearest neighbor graph: there are very few edges connecting nodes far away, and all pairs of close nodes are connected.

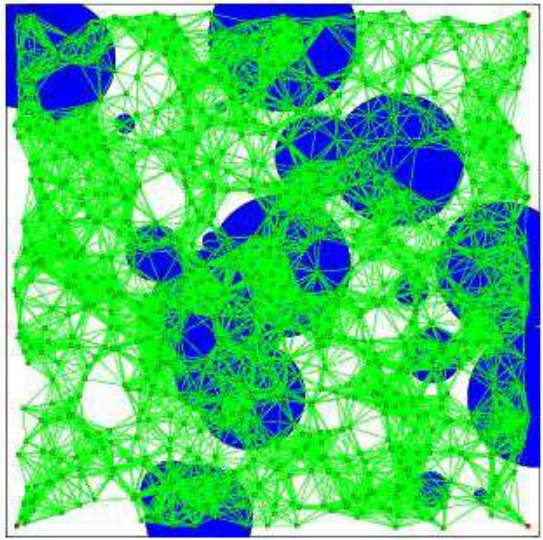

Fig. 3.  Boolean model with round clouds.

In this example, on each corner of the unit square, a beacon with known position (illustrated by a red square) is shown. Though only four beacons are given, based on the proximity graph, we still can give a reasonably good estimation of locations for most of the sensors. This result is discussed in Section V-E.

### B. Big clouds

The big clouds in this simulation were modeled by half planes (a realization is shown in blue on Figure 4) bounded by lines with isotropic orientation. Again, a given number ($k_N = 10$) of pairs of nodes with the largest value of the empirical cumulants were selected; a visual inspection indicates high similarity of this graph with the nearest neighboring graph.

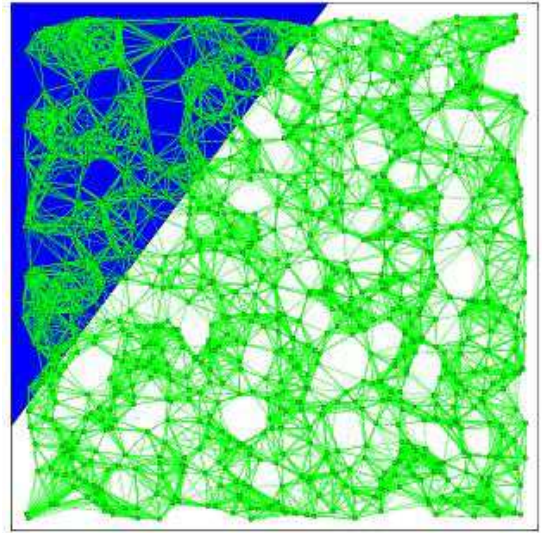

Fig. 4.  Random sets are formed by isotropic half-planes.

### C. Random walkers

We consider here $W = 10$ random walkers, which are monitored by sensors of a sensing radius $r = .13$. Yellow trajectories show part of the traces of the walkers.

Consider the lagged correlations, with lags equal to $\Delta = 2$. For each *sensor* we then select $k_N = 10$ neighbors with the best values of empirical cumulants.

One can see on Figure 5 that the quality of the proximity graph in this situation is almost the same as that in the previous two examples, even in the presence of the relatively slow convergence of the random fields defined by the walkers: to ensure the convergence of the empirical cumulants to their average values one needs at least the convergence of the (normalized) occupation measure to the uniform measure. While from Figure 5, it is evident that the occupation measure in the simulated example is still far away from the uniform measure (which can be seen from the scatter plot of cumulants illustrated in Figure 8 of Section V-D). Yet, even given this source of imperfection, the result can be reliably used to estimate the proximity of the nodes.

### D. Quality of distance approximation

A better feeling about the quality of the distance data recovery using the correlation between nodes can be gained from the scatter plots, which show the cumulant values versus the distances of pairs of sensors. The first two scatter plots, the round clouds model in Figure 6 and the large clouds model in Figure 7, show the results for *all pairs* of sensors.

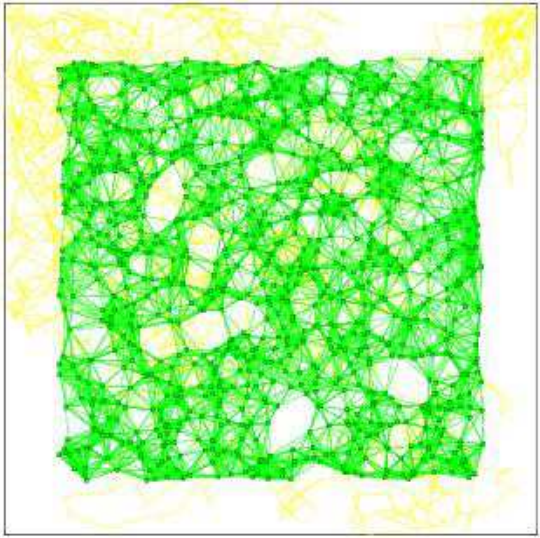

Fig. 5. Part of the traces of $W = 10$ random walkers (reflected at the boundary) are shown in yellow. The region in which the walkers move is larger than the region $A$ where the sensor network is deployed, to avoid irregularities at the boundary.

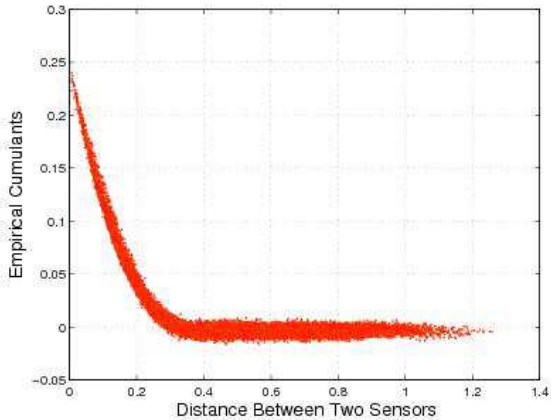

Fig. 6. Cumulant-distance scatter plot for the round clouds model. All pairs of points are shown.

For the random walkers model, the scatter plot (Figure 8) shows the cumulant-distance pairs with one of the nodes fixed. The heterogeneity of the occupation measure leads to significantly different ranges of the cumulants at different parts of the region. However, for any particular node, the cumulants can be very efficiently used for selecting the closest nodes, as the plot shows.

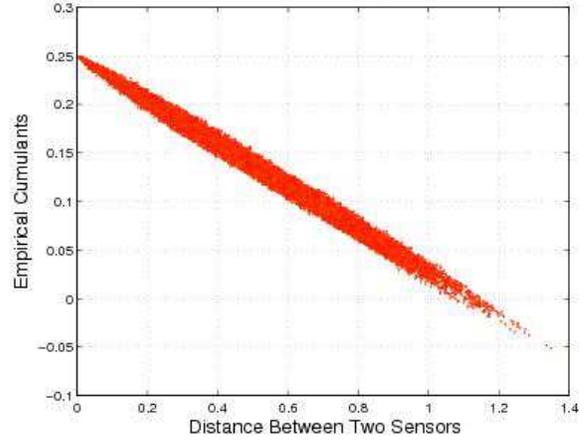

Fig. 7. Cumulant-distance scatter plot for the big clouds model. All pairs of points are shown.

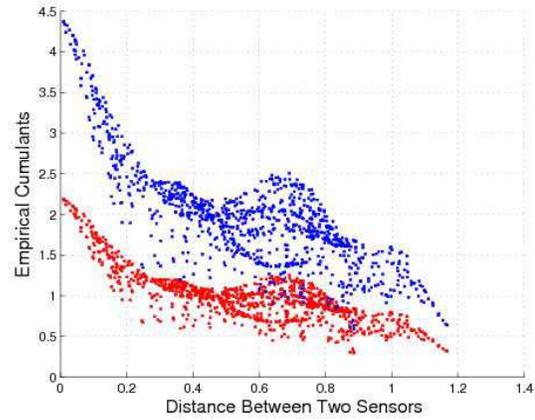

Fig. 8. Scatter plot of cumulants versus distances for all pairs of sensors with one fixed: blue dots show the data for one node in the region of high occupation density; red dots correspond to another node with low occupation density. The plots are visibly similar, exhibiting high reliability of cumulant estimator.

### E. End-to-end localization in anchoritic sensor networks

For the round clouds model, we used the cumulant-based proximity graph to approximate the internode distances and ultimately to reconstruct the positions of the sensors. Since the second step of reconstructing exact positions is not the main focus of this paper, although it is very important for localization, we choose to do it in a rather naive way. We compute the hop distances of the sensor of interest to the four beacons lying on the four corners using Dijkstra's algorithm. Assuming the hop distance is proportional to the real Euclidean distance (which is shown in Theorem 1), we can estimate the locations of all sensors, and the results are shown in Figure V-E. One can see that boundary effects (caused by the inefficient algorithm we used here) are rather significant, yet in the interior of the area the positions are recovered quite well.

Here we did not use any sophisticated machinery for the second step of the localization problem, and a more holistic approach would be to generate a Gibbs measure, the ensemble of $N$ nodes in $A$ whose distribution would consist with the empirical measurements. Sampling from this distribution would give the most probable positions of the sensors in the region $A$.

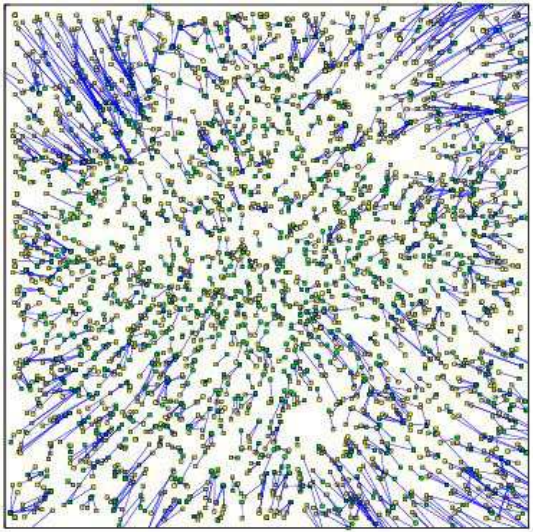

Fig. 9. Results of the end-to-end positions estimation in a anchoritic sensor network. Round clouds cumulants were used to generate the proximity graph. The actual node positions are shown as yellow squares connected to their estimated positions (green squares).

## VI. Conclusion

As demonstrated above, the cumulant evaluation technique allows one to address the localization problem in anchoritic sensor networks, where the chatter between the nodes is undesirable or infeasible. We would like to indicate several research directions prompted by the present study, which we hope to pursue:

1) We concentrated in this study on pairwise cumulants. This fits into the customary context of the existing approached to the localization problem in restricting attention to pairwise mutual positions of the nodes. Yet, the study of correlations allow one to gain much more information by considering the higher order cumulants. For example, in the big clouds models, the leading non-constant term in the $k$-order cumulant is proportional to the circumference of the convex hull of the nodes for which the cumulant is computed. Finding systematic ways to use this information is a challenging and promising task.

2) We approach the second step of the localization problem — finding the positions from the cumulant data — by generating the proximity graph. More holistic approach would be to generate a Gibbs measure, the ensemble of $N$ nodes in $A$ whose distribution would consist with the empirical measurements. Sampling from this distribution would give the most probable positions of the nodes in the region $A$.

## References

[1] K. Langendoen and N. Reijers, "Distributed localization in wireless sensor networks: a quantitative comparison," *Comput. Networks*, vol. 43, no. 4, pp. 499–518, 2003.

[2] C. Savarese, J. Rabaey, and J. Beutel, "Locationing in distributed ad-hoc wireless sensor networks," *ICASSP*, May 2001.

[3] R. L. Moses, D. Krishnamurthy, and R. Patterson, "An auto-calibration method for unattended ground sensors," *In ICASSP*, vol. 3, p. 2941¨C2944, May 2002.

[4] L. Dohert, K. Pister, and L. Ghaoui, "Convex position estimation in wireless sensor networks," *INFOCOM'01*, April 2001.

[5] A. Savvides, H. Park, and M. Srivastava, "The bits and flops of the N-hop multilateration primitive for node localization problems," *WSNA'02, Atlanta, Georgia, USA*, September 2002.

[6] S. Simic and S. Sastry, "Distributed localization in wireless ad hoc networks," 2002. [Online]. Available: citeseer.ist.psu.edu/simic01distributed.html

[7] P. Bergamo and G. Mazzini, "Localization in sensor networks with fading and mobility," *Personal, Indoor and Mobile Radio Communications (PIMRC'02)*, September 2002.

[8] T. He, C. Huang, B. Blum, J. stankovic, and T. Abdelzaher, "Range-free localization schemes for large scale sensor networks," *MobiCom'03, San Diego, CA, USA*, August 2003.

[9] D. Niculescu and B. Nath, "Localized positioning in ad hoc networks," *In Sensor Network Protocols and Applications, Anchorage, Alaska*, April 2003.

[10] N. Patwari and A. O. H. III, "Using proximity and quantized RSS for sensor localization in wireless networks," *WSNA'03.San Diego, CA,USA*, September 2003.

[11] Y. Shang, W. Ruml, Y. Zhang, and M. P. J. Fromherz, "Localization from mere connectivity," in *MobiHoc '03: Proceedings of the 4th ACM international symposium on Mobile ad hoc networking & computing*. New York, NY, USA: ACM Press, 2003, pp. 201–212.

[12] Y. Shang and W. Ruml, "Improved MDS-based localization," *Infocom 2004*, vol. 3, pp. 7–11, 2004 March.

[13] R. Nagpal, H. Shrobe, and J. Bachrach, "Organizing a global coordinate system from local information on an ad hoc sensor network," *2nd International Workshop on Information Processing in Sensor Networks (IPSN 03)*, April 2003.

[14] J. Albowicz, A. Chen, and L. Zhang, "Recursive position estimation in sensor networks," *Ninth International Conference on Network Protocols*, November 2001.

[15] D. Niculescu and B. Nath, "Ad hoc positioning system (APS)," *Global Telecommunications Conference 2001. GLOBECOM '01. IEEE*, vol. 5, November 2001.

[16] R. Stoleru, T. He, J. A. Stankovic, and D. Luebke, "High-accuarcy,low-cost localization system for wireless sensor networks," *In Third ACM Conference on Embedded Networked Sensor Systems (SenSys2005)*, November 2005.

[17] M. Stojanovic, "Acoustic (underwater) communications," *entry in Encyclopedia of Telecommunications, John G. Proakis, Ed., John Wiley & Sons*, 2003.

[18] N. Bulusu, J. Heidemann, and D. Estrin, "GPS-less low cost outdoor localization for very small devices," *IEEE Personal Communications Magazine*, vol. 7, no. 5, pp. 28–34, October 2000.

[19] N. B. Priyantha, A. K. L. Miu, H. Balakrishnan, and S. Teller, "The cricket compass for context-aware mobile applications," *Proc. of the 6th ACM MOBICOM Conf., Rome, Italy*, July 2001.

[20] L. Girod, V. Bychkovskiy, J. Elson, and D. Estrin, "Locating tiny sensors in time and space: A case study," *In Proceedings of the International Conference on Computer Design (ICCD2002), Freiburg, Germany*, September 2002.

[21] S. Funiak, C. Guestrin, M. Paskin, and R. Sukthankar, "Distributed localization of networked cameras," in *IPSN '06: Proceedings of the fifth international conference on Information processing in sensor networks*. New York, NY, USA: ACM Press, 2006, pp. 34–42.

[22] O. Kallenberg, *Foundations of Modern Probability*. Springer Series in Statistics. Probability and Its Applications, October 1997.

[23] D. Ruelle, *Statistical Mechanics, Rigorous results*. Benjamin, 1969.

[24] J. M. D. Stoyan, W.S. Kendall, *Stochastic geometry and its applications*. Wiley, 1987.